\documentclass[showpacs,preprintnumbers,amsmath,amssymb,twocolumn]{revtex4-1}
\usepackage{graphicx}
\usepackage{tabularx}
\usepackage{booktabs}
\usepackage{mathtools}
\usepackage{enumerate}
\usepackage{epstopdf}
\usepackage[colorlinks=true,allcolors=blue]{hyperref}
\usepackage[all]{hypcap}
\usepackage{float}
\usepackage[usenames,dvipsnames]{color}
\usepackage{diagbox}

\begin{document}
   \title{Time Asymmetry of Cosmic Background Evolution in Loop Quantum Cosmology}
   \author{Wen-Hsuan Lucky Chang}
%     \email{f00222018@ntu.edu.tw}
%     \affiliation{%
%Department of Physics, National Taiwan University, 
%Taipei 10617, Taiwan
%}%
   \author{Jiun-Huei Proty Wu}
%     \email{jhpw@phys.ntu.edu.tw}
     \affiliation{%
Department of Physics, Institute of Astrophysics, and Center for Theoretical Physics, National Taiwan University, Taipei 10617, Taiwan
}%

%	\date{\today}

\begin{abstract}
We discuss the asymmetry of cosmic background evolution in time with respect to the quantum bounce in the Loop Quantum Cosmology (LQC), employing the value of scalar field at the bounce $\phi_{\rm B}$.
We use the Chaotic and the $R^2$ potentials to demonstrate that a possible deflation before the bounce may counteract the inflation that is needed for resolving the cosmological conundrums, 
so a certain level of time asymmetry is required for the models in LQC.
This $\phi_{\rm B}$ is model dependent and closely related to the amounts of deflation and inflation, so we may use observations to confine $\phi_{\rm B}$ and thus the model parameters.
With further studies this formalism should be useful in providing an observational testbed for the LQC models.
\end{abstract}

\maketitle

% -----------------------------------------------------------------------

\section{Introduction}\label{sec1}

Singularity at the beginning of spacetime is a long-standing problem in cosmology \cite{Hawking1970}. One solution is to consider the Loop Quantum Cosmology (LQC), which is a theory of Loop Quantum Gravity (LQG) simplified with the cosmological principle \cite{Bojowald2002a}. It employs the Friedmann-Robertson-Walker (FRW) model with quantum corrections. The extra terms involve a scalar field to resolve the singularity problem with a quantum bounce \cite{Bojowald2001a}. In turn this allows for the existence of a  `parent universe' \cite{Ashtekar2006b,Ashtekar2006c,Ashtekar2006d}.

To evolve the scale factor under this context,
the quantum corrected Friedmann equation \cite{Bojowald2005,Vandersloot2005} was derived with the Hamiltonian formulation in a semi-classical approach   \cite{Bojowald2001b}. Two major types of quantum corrections are the holonomy \cite{Singh2005,Chiou2009,Chiou2009a} and the inverse volume \cite{Bojowald2001c}.
The Hamiltonian involves the connection variables (known as the Ashtekar variables in LQC), whose equations of motion can be obtained by calculating their Poisson brackets. Because these connection variables are actually functions of the scale factor and the Hubble parameter, we could eventually obtain the evolution equation of the scale factor (see Ref.~\cite{Banerjee2012} for introductory review).

Within the LQC framework, inflation occurs naturally after the quantum bounce due to the existence of a scalar field \cite{Bojowald2002}, so that the cosmological conundrums can be resolved in the conventional way \cite{Guth1981}. 
Before the quantum bounce this scalar field may also generate a period of damped contraction called `deflation'.
The amount of deflation and that of inflation may differ and one key is the potential to kinetic energy ratio (PKR) of the scalar field at the quantum bounce.
Here we shall directly employ a more intuitive quantity $\phi_{\rm B}$, the $\phi$ value at the quantum bounce, to study the asymmetry between deflation and inflation.

This paper investigates in details the dependence of cosmic time asymmetry on $\phi_{\rm B}$. 
The two inflationary models considered here are the Chaotic potential (a commonly chosen simple model) and the $R^2$ potential (a realistic model to date \cite{Martin2014}).

Here is the structure of this paper.
First we lay out our convention of LQC in Section \ref{sec2},
where Section \ref{ch2.3.1} defines the Hamiltonian formalism, with its quantum corrections presented in Section \ref{ch2.3.2}. Section \ref{sec3} investigates the time asymmetry in cosmic evolution, in particular employing the $\phi$ value at the quantum bounce. Section \ref{sec4} discusses the possible deflation  and its impact. We conclude our work in Section \ref{sec6}.
The units of all physical quantities in this paper are normalized to the Planckian units ($c=G=\hbar=k_{\rm B}\equiv1$) unless otherwise labeled. The curvature constant is also presumed to be zero, which is consistent with the current observational results.

% -----------------------------------------------------------------------

\section{Cosmic Dynamics}\label{sec2}
\subsection{Hamiltonian formalism}\label{ch2.3.1}

To have a good handle on the quantum mechanical properties of the early universe, we employ the Arnowitt-Deser-Misner (ADM) approach.
The Hamiltonian of spacetime is
\begin{align}\label{eq2.14}
H_{\rm grav} = -\frac{3}{8\pi\gamma^2} c^2 \sqrt{p},
\end{align}
where $p$ and $c$ (not the speed of light) are the connection variables, which are related to the scale factor $a$ and the Hubble parameter $H$ as \cite{Bojowald2005}
\begin{align}
|p| & = \frac{1}{4}a^2, \label{eq.p}\\
c & = \frac{1}{2}\gamma aH, \label{eq.c}
\end{align}
and satisfy the canonical relation \cite{Bojowald2005}
\begin{align}\label{eq2.16}
\left[c,p\right]_{\rm PB} = \frac{8\pi\gamma}{3}.
\end{align}
The subscript `PB' denotes that the calculation rule follows Possion Bracket rather than the commutation.
The Barbero-Immirzi parameter \cite{BarberoG.1995,Immirzi1997} $\gamma=\log(3)/\sqrt{2}\pi$ can be obtained from the computation of black hole entropy \cite{Meissner2004a}. 
In Eqs.~\eqref{eq.p} and \eqref{eq.c} we have dropped the curvature parameter of the FRW model and chosen the coordinate length of the finite-sized cubic cell in LQG to be unity. It is obvious that the energy density of the spacetime 
\begin{align}\label{eq.rho}
	\rho_{\rm grav}=p^{-3/2}H_{\rm grav}
\end{align}
is unbounded when the size of the universe goes to zero ($a\rightarrow 0$).

On the other hand, the Hamiltonian of the inflaton, which is the only content that matters during a single-field inflation, is
\begin{align}\label{eq2.17}
H_\phi = \frac{\pi_\phi^2}{2p^{3/2}} + p^{3/2}V(\phi),
\end{align}
where the scalar field $\phi$ and its conjugate momentum $\pi_\phi$ satisfy the canonical relation \cite{Chiou2009}
\begin{align}\label{eq2.18}
\left[\phi,\pi_\phi\right]_{\rm PB} = 1.
\end{align}
General Relativity (GR) then requires that
the total Hamiltonian must be zero at all times:
\begin{align}\label{eq2.19}
H_{\rm tot} = H_{\rm grav} + H_{\phi} = 0.
\end{align}
This is the Hamiltonian constraint, which is commonly used in solving the Einstein equations numerically \cite{53684}.
Consequently, the equations of motion that describe the dynamics of the universe are
\begin{align}\label{eq2.20}
\frac{dq}{dt} &= [q,H_{\rm tot}]_{\rm PB},
\end{align}
where $q$ represents $p$, $c$, $\phi$, or $\pi_\phi$ \cite{Grain2010}. This set of equations are equivalent to the Friedmann equation and the fluid equation.

\subsection{Holonomy corrections}\label{ch2.3.2}

For the quantum corrections in the above formalism,
we adopt a semi-classical approach in LQC \cite{Bojowald2001b}.
The $n$th-order holonomized connection variable $c_h^{(n)}$ is defined as \cite{Chiou2009}
\begin{align}\label{eq2.21}
c_h^{(n)} \equiv \frac{1}{\bar{\mu}}\sum_{k=0}^n \frac{(2k)!}{2^{2k}(k!)^2(2k+1)}(\sin\bar{\mu}c)^{2k+1},
\end{align}
where $\bar{\mu}=\sqrt{\Delta/p}$ is the discreteness variable with $\Delta=2\sqrt{3}\pi\gamma$ being the standard choice of the area gap in the full theory of LQG \cite{Ashtekar2006b}. One key feature in LQC is that the connection variable in the standard cosmology has to be replaced by holonomies. Thus the Hamiltonian of spacetime with the holonomy correction up to the $n$th-order is \cite{Chiou}
\begin{align}\label{eq2.22}
H^{(n)}_{\rm grav,\bar{\mu}} = -\frac{3}{8\pi\gamma^2} (c_h^{(n)})^2 \sqrt{p}.
\end{align}
Finally the new Hamiltonian constraint is \cite{Chiou,Mielczarek2010}
\begin{align}\label{eq2.23}
H^{(n)}_{\bar{\mu}} = H^{(n)}_{\rm grav,\bar{\mu}} + H_\phi = 0.
\end{align}
We can apply this to the semi-classical approach as what was done in GR \cite{Thiemann1998}.

With such quantum corrections, it is obvious that the energy density of the spacetime $\rho_{\rm grav}$ is always finite. The extreme values appear when $\bar{\mu}c$ equals $0$, $\pi/2$, or its multiples. When $\bar{\mu}c=\pi/2$, the Hamiltonian $H^{(n)}_{\rm grav,\mu}$ reaches its minimum and thus $H_\phi$ reaches its maximum. The maximal energy density of the inflton $\rho_\phi=p^{-3/2}H_\phi$ is called the `critical energy density' and is related to the holonomies as \cite{Chiou2009}
\begin{align}\label{eq2.24}
\rho_{\rm c}^{(n)} &= \frac{\sqrt{3}m_{\rm pl}^4}{16\pi^2\gamma^3} \left[\sum\limits_{k=0}^n \frac{(2k)!}{2^{2k}(k!)^2(2k+1)}\right]^2,
\end{align}
which is confined between $\rho_{\rm c}^{(0)}\simeq 0.82m_{\rm pl}^4$ and $\rho_{\rm c}^{(\infty)}\simeq 2.02m_{\rm pl}^4$.
We note that the standard cosmology is recovered ($c_h^{(n)} \rightarrow c$) when $\bar{\mu}c\rightarrow 0$ (that is, when $p\gg 1$). This indicates that the quantum effects are important only when the universe is tiny ($p\sim\Delta$).

Consequently the equations of motion can be obtained as
\begin{align}\label{eq2.25}
\frac{dq}{dt} &= [q,H^{(n)}_{\bar{\mu}}]_{\rm PB},
\end{align}
which are equivalent to the Friedmann equation and the fluid equation with quantum corrections \cite{Bojowald2005,Vandersloot2005}.
According to the literatures \cite{Chiou2009a,Chiou2009}, the higher-order effects on the cosmic background are distinguishable but secondary.

At the end of this section, we note that we choose the lapse function as one in this paper. The time parameter ‘$t$’ therefore corresponds to the coordinate time of a FRW metric in the classical regime.

% -----------------------------------------------------------------------

\section{Cosmic time asymmetry}\label{sec3}
\subsection{The Bouncing Scenario}

\begin{figure}[t!]
	\centering
	\includegraphics[width=0.48\textwidth]{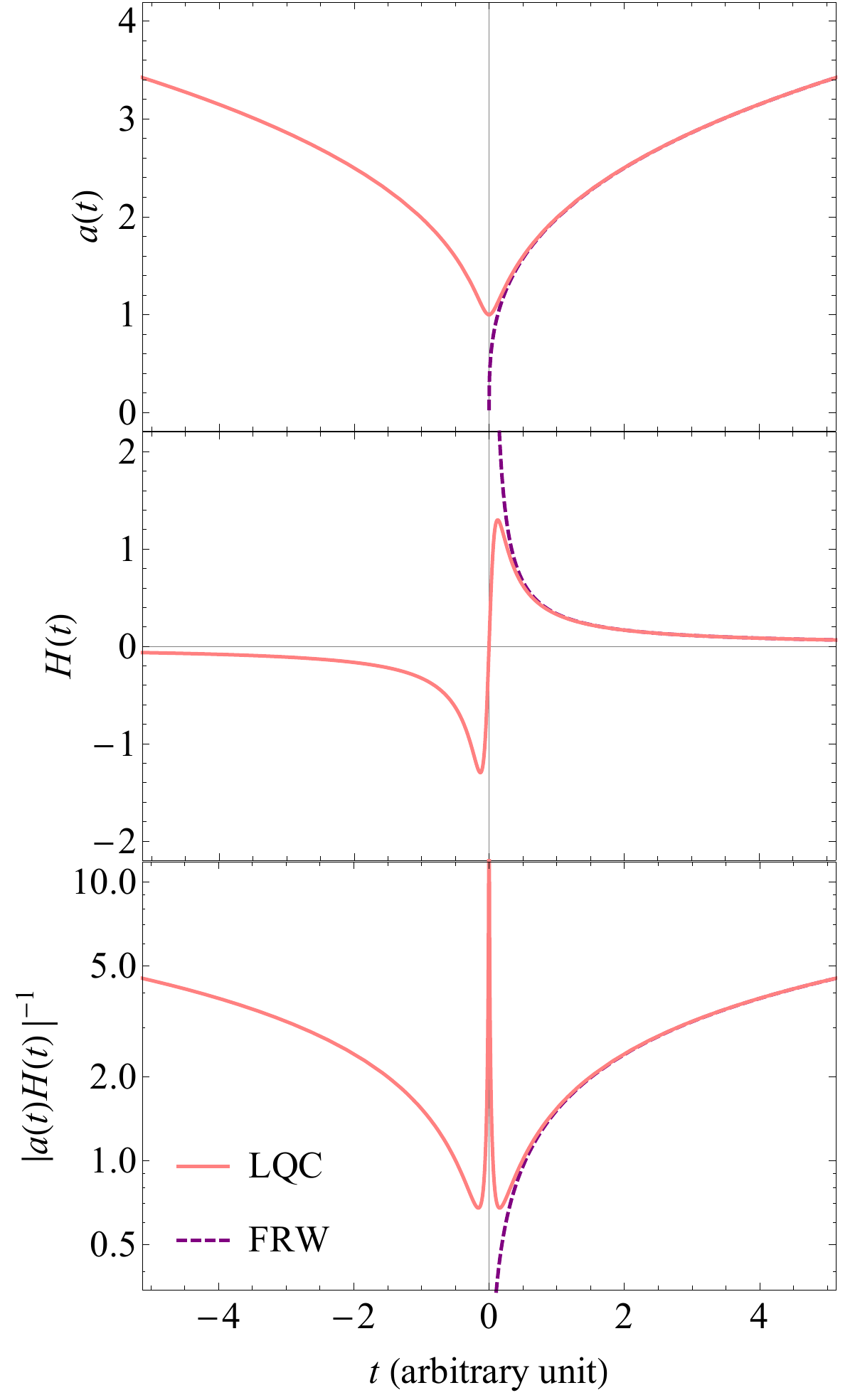}
	\caption{\label{fig2.1}The time evolution of the scale factor (top), the Hubble parameter (middle), and the comoving Hubble radius (bottom). We consider $V(\phi)=0$ in this figure for simply demonstrating the quantum bounce.}
\end{figure}

As we have seen in the previous section, the energy density of the scalar field now has a maximum $\rho_{\rm c}^{(n)}$ (when $\bar{\mu}c=\pi/2$) and thus avoids the singularity. 
To see how this is manifested in the behavior of cosmic expansion,
we can use the equations of motion to first obtain the Hubble parameter  as \cite{Chiou}
\begin{align}\label{eq2.26}
H &= \frac{\dot{a}}{a} = \frac{2}{p}[p,H^{(n)}_{\bar{\mu}}]_{\rm PB} \nonumber \\
&= \frac{4}{\gamma\sqrt{p}} \cos (\bar{\mu}c) \mathcal{O} _n(\bar{\mu}c) c_h^{(n)},
\end{align}
where
\begin{align}\label{eq2.27}
\mathcal{O} _n(\bar{\mu}c) \equiv \sum _{k=0}^{n} \frac{(2k)!}{2^{2k}(k!)^2}
(\sin\bar{\mu}c)^{2k}.
\end{align}
The solid curves in Fig.~\ref{fig2.1} are the numerical solutions of the scale factor $a$, the Hubble parameter $H$, and the comoving Hubble radius $|H^{-1}/a|$, as functions of time~$t$.
The scale factor $a(t)$ is normalized to unity at $t=0$.
It shows that the universe contracts before the quantum bounce and expands after the bounce, with a turning point of $a(0)=1$ corresponding to $\bar{\mu}c=\pi/2$.
We refer to the epoch before the bounce as the `parent universe'.

For the Hubble parameter, it changes its sign at the bounce.
The fact that $H(0)=0$ indicates that the comoving Hubble radius $|H^{-1}/a|$ diverges to infinity at the bounce. This means that the quantum effects are extremely strong such that the whole universe is in causal contact at the bounce. 

The dashed curves in Fig.~\ref{fig2.1} are the results in the standard cosmology, without the quantum corrections.
In this case the universe starts from singularity at $t=0$, without causal connections at all because the comoving Hubble radius is zero at this time. 

While the solid curves show symmetry in time with respect to the quantum bounce at $t=0$,
such symmetry may be broken in general cases.
According to Eqs.~\eqref{eq2.17}, \eqref{eq2.22}, \eqref{eq2.23},  \eqref{eq2.24}, and \eqref{eq2.25}, we have
\begin{align}\label{eq2.28}
\frac{1}{2}\dot{\phi}^2 + V(\phi) = \rho_{\rm c}^{(n)},
\end{align}
which is a constant for given $n$.
Thus the PKR of the scalar field at the bounce is a free parameter so we may define a `bouncing phase' as
\begin{align}\label{eq2.29}
\theta_{\rm B} = \tan^{-1} \frac{\sqrt{2V(\phi)}}{\dot{\phi}}.
\end{align}
For the cases where $V(\phi)$ is an even or odd function in $\phi$, $\theta_{\rm B}$ determines the level of time asymmetry in the cosmic background dynamics. 
The case $\theta_{\rm B}=0$ (and thus $\phi=0$ at bounce for the scalar-field potentials considered in this paper) corresponds to a time symmetry with respect to $t=0$;
other cases lead to time asymmetry. 
For the cases where $V(\phi)$ is not symmetric in $\phi$, the cosmic background dynamics is always asymmetric with respect to the bounce. 
Ref.~\cite{Mielczarek2010} studied a special kind of asymmetric cases called the `shark-fin type', which provides a relatively large number of $e$-foldings in the inflation after quantum bounce.

\subsection{Realistic Scalar Models}

Eq.~\eqref{eq2.29}, however, has a limit that $V(\phi)$ must stay non-negative, and thus cannot be applied to a general potential. Also, the PKR does not have one-to-one correspondence to the time symmetry. Due to these reasons we consider directly the $\phi$ value at the quantum bounce, labeled as $\phi_{\rm B}$, as a free parameter that quantifies the symmetry. Because the number of $e$-foldings in inflation depends on the value of $\phi$, the value of $\phi_{\rm B}$ is more apparently related to the intrinsic properties of an inflationary model than $\theta_{\rm B}$. For scalar potentials symmetric in $\phi$, the case $\phi_{\rm B}=0$ corresponds to a time-symmetric case; in a time-asymmetric case, the $\phi$ value at the end of deflation would differ from the beginning of inflation leading to a non-zero $\phi_{\rm B}$.

\begin{figure}[t!]
	\centering
	\includegraphics[width=0.48\textwidth]{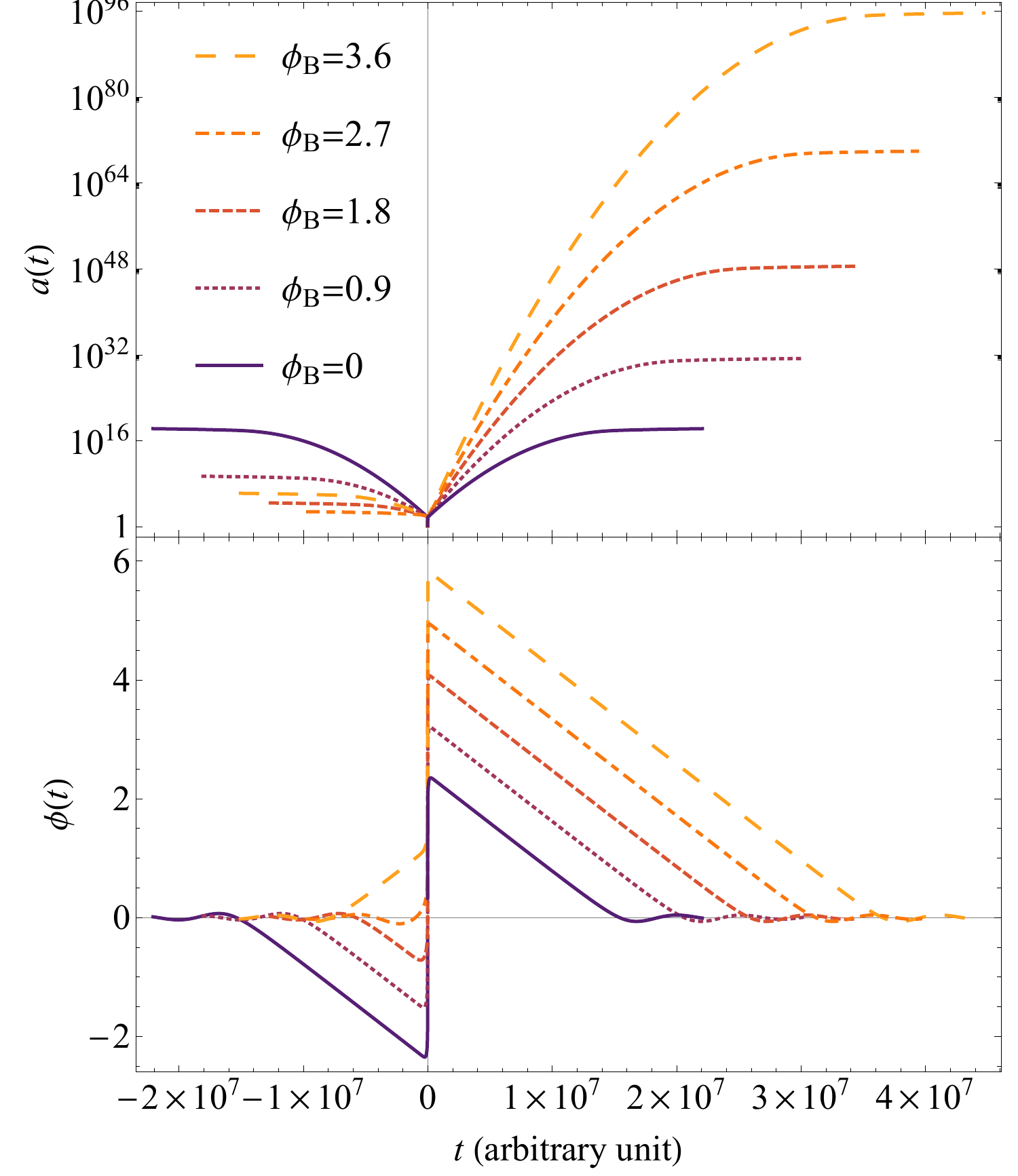}
	\caption{\label{fig2.4}The scale factor (upper panel) and the scalar field (lower panel) as functions of time at different values of $\phi_{\rm B}$, for Chaotic potential.}
\end{figure}

Given this new parameter $\phi_{\rm B}$, we first consider the Chaotic inflation 
\begin{align}\label{eq2.33}
V(\phi) = \frac{1}{2}m_\phi^2\phi^2.
\end{align}
Fig.~\ref{fig2.4} shows the scale factor and the scalar field as functions of time, at different values of $\phi_{\rm B}$.
We have considered the zeroth-order holonomy correction ($n=0$) and chosen the inflaton mass $m_\phi=10^{-6}$ in deriving the results in this figure. 
It is clear that $\phi_{\rm B}=0$ corresponds to a time-symmetric case, while
a larger $\phi_{\rm B}$ corresponds to a larger initial $\phi$ at the beginning of inflation, leading to a larger number of $e$-foldings. In addition, the amount of deflation is less when $\phi_{\rm B}$ is larger. The shark-fin type in Ref.~\cite{Mielczarek2010} corresponds to our case with $\phi_{\rm B} \approx 2.7$, where the time asymmetry is about the largest.
Table~\ref{tab2.1} shows the number of $e$-foldings for Chaotic inflation with various  $\phi_{\rm B}$ and $m_\phi$. 

\begin{table}[t!]
\centering
\begin{tabular}{|c|m{3em}|m{3em}|m{3em}|m{3em}|m{3em}|m{0em}}
\cline{1-6}
\diagbox{$m_\phi$}{$\phi_{\rm B}$} & $0$ & $0.9$ & $1.8$ & $2.7$ & $3.6$ & \\
\cline{1-6}
$10^{-4}$ & $18.5$ & $33.0$ & $52.2$ & $76.0$ & $105$ & \\ [3pt]
\cline{1-6}
$10^{-6}$ & $36.3$ & $66.6$ & $107$ & $157$ & $217$ & \\ [3pt]
\cline{1-6}
$10^{-8}$ & $60.6$ & $113$ & $183$ & $270$ & $374$ & \\ [3pt]
\cline{1-6}
$10^{-10}$ & $91.7$ & $173$ & $280$ & $414$ & $575$ & \\ [3pt]
\cline{1-6}
\end{tabular}
\caption{The number of $e$-foldings for inflation with Chaotic potential.}
\label{tab2.1}
\end{table}
\begin{table}[t!]
\centering
\begin{tabular}{|c|m{3em}|m{3em}|m{3em}|m{3em}|m{3em}|m{0em}}
\cline{1-6}
\diagbox{$m_\phi$}{$\phi_{\rm B}$} & $0$ & $0.9$ & $1.8$ & $2.7$ & $3.6$ & \\
\cline{1-6}
$10^{-4}$ & $18.5$ & $8.45$ & $2.56$ & $0.33$ & $3.66$ & \\ [3pt]
\cline{1-6}
$10^{-6}$ & $36.3$ & $15.6$ & $3.97$ & $0.25$ & $8.84$ & \\ [3pt]
\cline{1-6}
$10^{-8}$ & $60.6$ & $25.1$ & $5.64$ & $0.19$ & $16.5$ & \\ [3pt]
\cline{1-6}
$10^{-10}$ & $91.7$ & $36.9$ & $7.55$ & $0.17$ & $26.2$ & \\ [3pt]
\cline{1-6}
\end{tabular}
\caption{The number of $e$-foldings for deflation with Chaotic potential.}
\label{tab2.2}
\end{table}

\begin{figure}[t]
	\centering
	\includegraphics[width=0.48\textwidth]{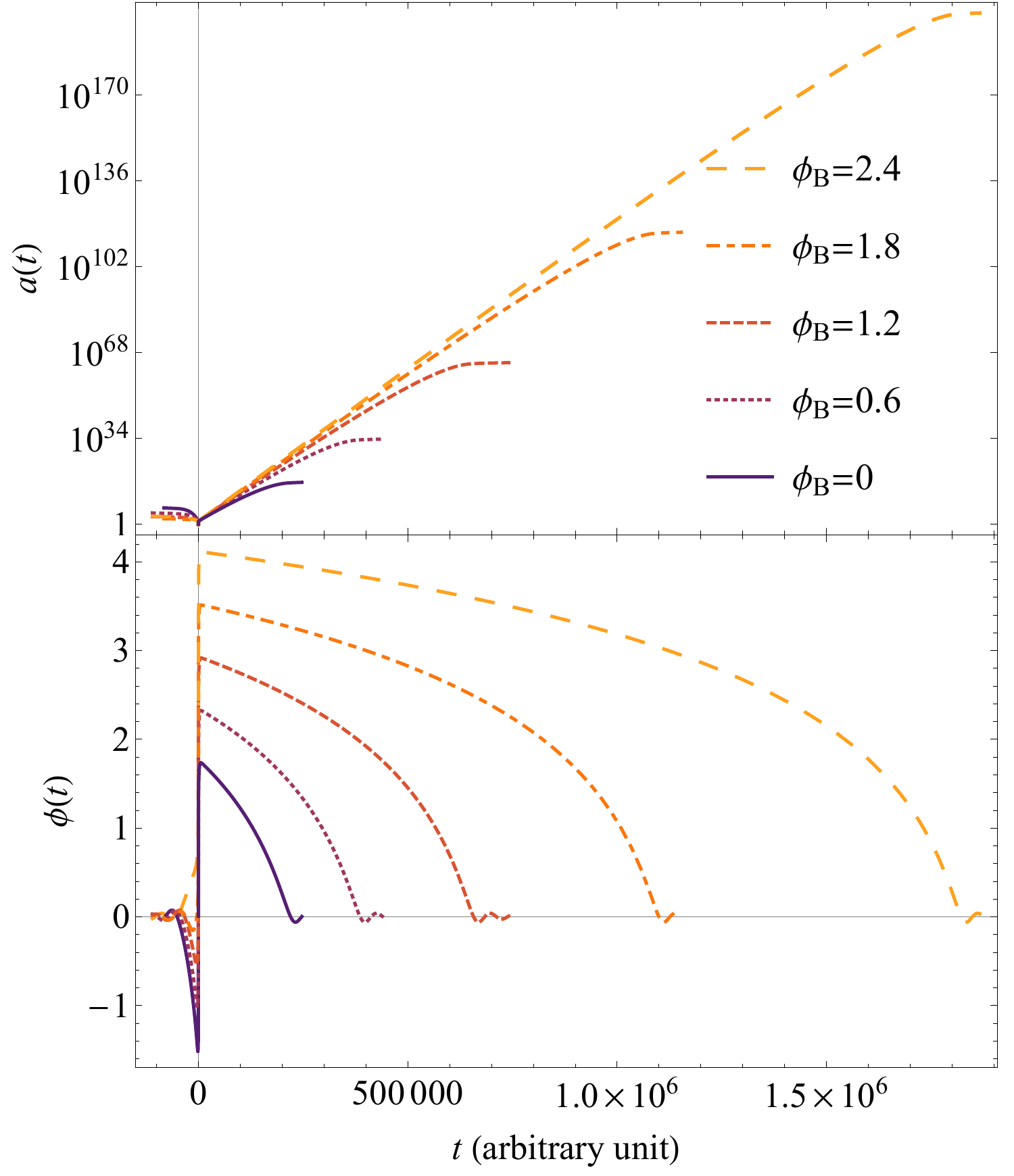}
	\caption{\label{fig2.5}The same as Fig.~\ref{fig2.4} for the $R^2$ potential.}
\end{figure}

Next we consider the $R^2$ inflation
\begin{align}\label{eq2.34}
V(\phi) = m_{\rm H}^4 \left( 1 - e^{-\sqrt{\frac{2}{3}}\phi} \right),
\end{align}
where $m_{\rm H}$ is the inflaton mass, which is normally denoted as $\Lambda$ in literature. Here the subscript `H' stands for the Higgs-like particle.
To clarify, this $R^2$ potential is not a quantum field in Starobinsky gravity but a classical field in GR.
The resulting time evolutions of the scale factor and the scalar field at different values of $\phi_{\rm B}$ are presented in Fig.~\ref{fig2.5}, where we have used $n=0$ and $m_{\rm H}=10^{-2}$.
Unlike the Chaotic inflation, here we see no case with time symmetry simply because the $R^2$ potential is not symmetric in $\phi$. 

These results also indicate that the cosmological inflation occurs naturally after the quantum bounce, with its initial condition unambiguously and naturally determined rather than manipulatively designed. This fact was previously studied for both time-symmetric background \cite{Chiou} and time-asymmetric background \cite{Mielczarek2010}. The four conditions required for solving the four coupled equations of motion are the Hamiltonian constraint $H^{(n)}_{\bar{\mu}}=0$, the turning point condition $\bar{\mu}c=\pi/2$, the value of $\phi_{\rm B}$, and the normalization of the scale factor $a$.

% -----------------------------------------------------------------------

\section{Cosmological Deflation}\label{sec4}
\subsection{Quantifying Deflation}\label{sec4.1}

When we look into the epoch right before the quantum bounce, the scalar field may induce a damped contraction of the space, which we call the `cosmological deflation'. During the deflation, we have
\begin{align}\label{eq2.35}
\dot{a}<0, \quad \ddot{a}>0.
\end{align}
In contrast to inflation, the comoving Hubble radius grows with time during deflation. In other words, the size of causally contacted region is increasing. In addition, the energy densities and thus the perturbations are increasing. All these may counteract the inflationary effects that we need for resolving the cosmological conundrums, so a scenario with comparably less deflation is in general needed. This in turn requires asymmetry in time with respect to the quantum bounce.

\begin{figure}[t]
	\centering
	\includegraphics[width=0.48\textwidth]{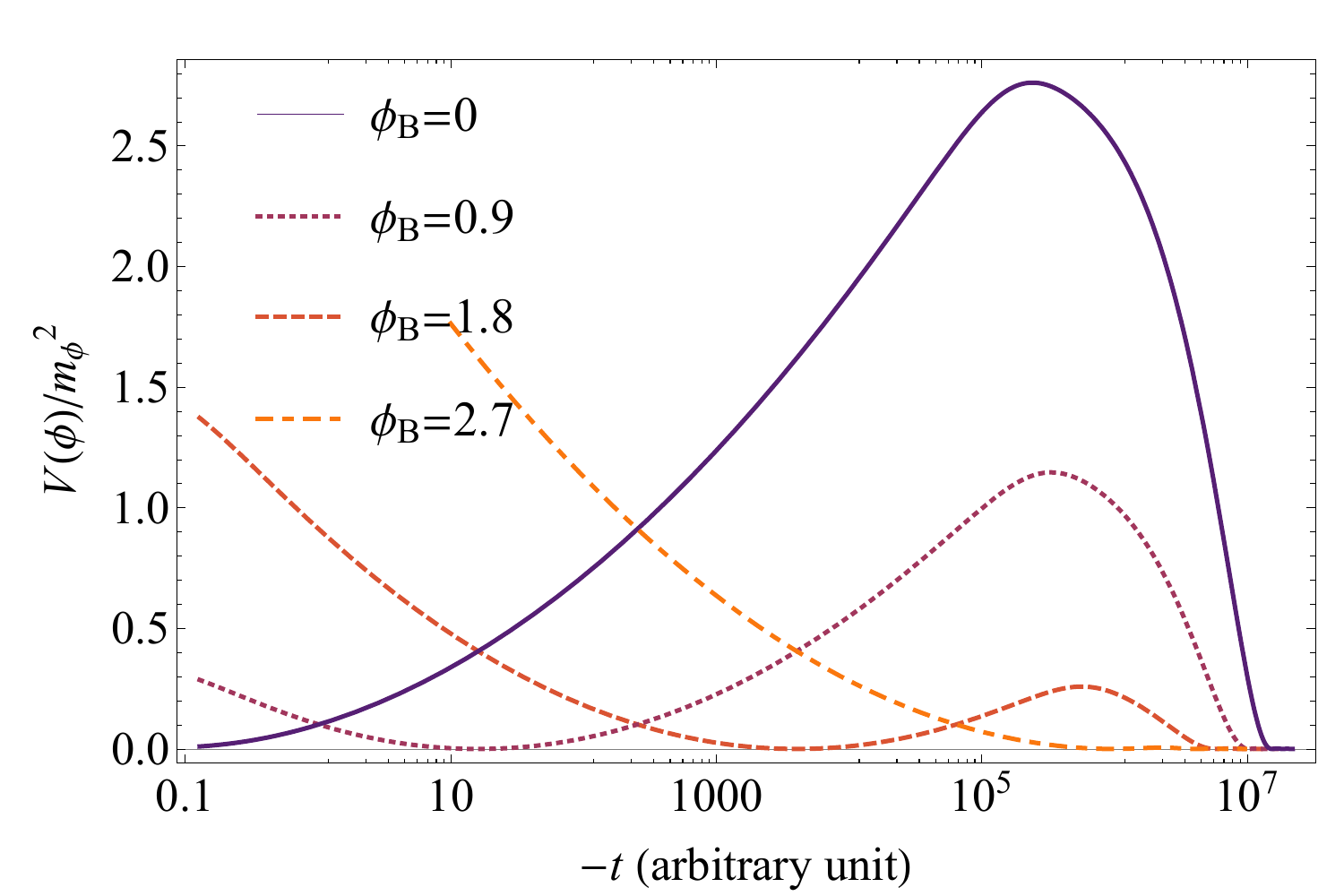}
	\caption{\label{fig2.6}Evolution of the normalized Chaotic potential before quantum bounce. The time goes leftwards in the figure, with the bounce as its origin.}
\end{figure}

Most of the formalisms used for the study of inflation are equally useful for the study of deflation, for example, the slow-roll approximation.
Fig.~\ref{fig2.6} shows how the Chaotic potential evolves with time before the quantum bounce. Deflation takes place when the slope is small and thus near the peaks of the curves in the figure.
For deflation we define the number of $e$-foldings similar to that of the inflation as
\begin{align}\label{eq2.36}
N_e^{\rm D} \equiv \ln \left( \frac{a_{\rm b}^{\rm D}}{a_{\rm e}^{\rm D}} \right),
\end{align}
where $a_{\rm b}^{\rm D}$ and $a_{\rm e}^{\rm D}$ are the scale factors at the beginning and the end of deflation respectively.
For a Chaotic potential under the slow-roll approximations, with $\phi_{\rm e}^{\rm D}$ the $\phi$ value at the end of deflation, this reduces to \cite{Jiun-HueiProtyWu1996}
\begin{align}\label{eq2.37}
N_e^{\rm D} \simeq 2\pi \left(\phi_{\rm e}^{\rm D}\right)^2 - \frac{1}{2}
= 4\pi \frac{V(\phi_{\rm e}^{\rm D})}{m_{\phi}^2} - \frac{1}{2}.
\end{align}
Combining this with Fig.~\ref{fig2.6}, we see the dependence of $N_e^{\rm D}$ on $\phi_{\rm B}$.
The dependence of $N_e^{\rm D}$ on $m_\phi$ is implicit as $a_{\rm b}^{\rm D}$ and $a_{\rm e}^{\rm D}$ are dependent on $m_\phi$.
Table~\ref{tab2.2} shows the dependence of $N_e^{\rm D}$ on some discrete values of $m_\phi$ and $\phi_{\rm B}$.
We see that for a fixed value of $m_\phi$ the case $\phi_{\rm B}=2.7$ always gives the least amount of deflation, as we can also see in Fig.~\ref{fig2.6} when combined with Eq.~\eqref{eq2.37}.
A comparison between Tables~\ref{tab2.1} and \ref{tab2.2} also shows that the case $\phi_{\rm B}=0$ has the same amount of inflation and deflation so their effects are expected to be reciprocally canceled out.
This is the time-symmetric case. Such scenarios are of less our interest because the cosmological conundrums revive here.
In the following we shall discuss the circumstances where such cancellation can be minimized.

\subsection{Minimizing Deflation}\label{ch2.3.4}

\begin{figure}[h]
	\centering
	\includegraphics[width=0.48\textwidth]{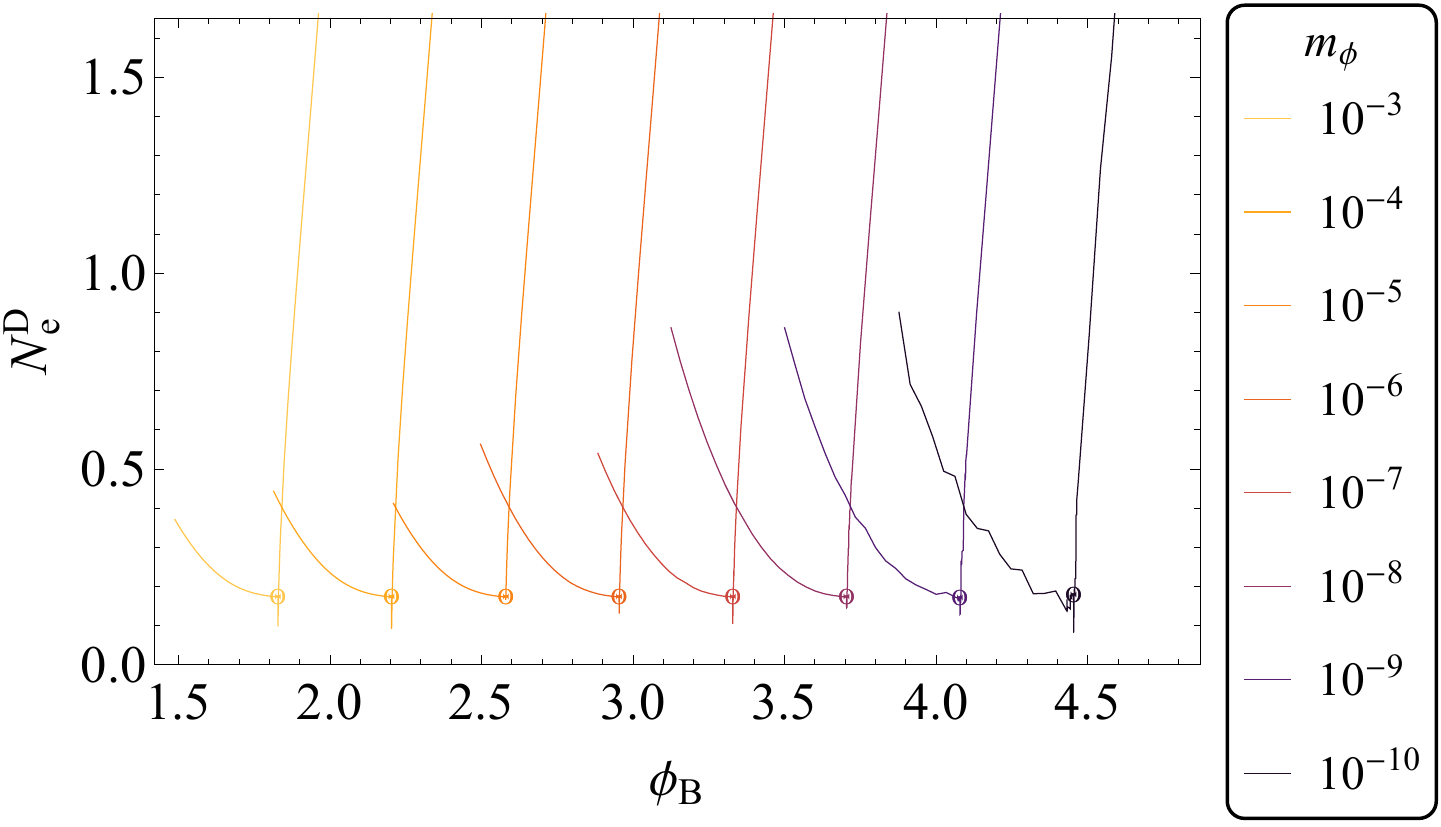}
	\caption{\label{fig2.7}The number of $e$-foldings $N_e^{\rm D}$ for Chaotic deflation as functions of $\phi_{\rm B}$ for different $m_\phi$.}
\end{figure}

We first numerically determine how $N_e^{\rm D}$ depends on $\phi_{\rm B}$.
For the Chaotic potential, Fig.~\ref{fig2.7} shows the $N_e^{\rm D}$ as a function of $\phi_{\rm B}$ at different but fixed values of $m_\phi$.
We use $\phi_{\rm crit}$ to denote the value of $\phi_{\rm B}$ at which the minimum $N_e^{\rm D}$ occurs in a curve.
It is interesting to note that the minimum values of $N_e^{\rm D}$ in all cases are about the same, $0.17$.
We also find that $\phi_{\rm crit}$ increases with $m_\phi$.

\begin{table}[h]
	\centering
	\begin{tabular}{|m{2.4em}|m{2.4em}|m{2.4em}|m{2.4em}|m{2.4em}|m{2.4em}|m{2.4em}|m{2.4em}|m{2.4em}|m{0em}}
		\cline{1-9}
		$m_\phi$ & $10^{-3}$ & $10^{-4}$ & $10^{-5}$ & $10^{-6}$ & $10^{-7}$ & $10^{-8}$ & $10^{-9}$ & $10^{-10}$ & \\ [5pt]
		\cline{1-9}
		$\phi_{\rm crit}$ & $1.83$ & $2.20$ & $2.58$ & $2.95$ & $3.33$ & $3.70$ & $4.08$ & $4.45$ & \\ [5pt]
		\cline{1-9}
	\end{tabular}
	\caption{The values of $\phi_{\rm crit}$ for different $m_\phi$ in Chaotic deflation. They correspond to the minima in Fig.~\ref{fig2.7}.}
	\label{tab2.3}
\end{table}

Table~\ref{tab2.3} summarizes the $\phi_{\rm crit}$ for different $m_\phi$. Here we surprisingly find that $\phi_{\rm crit}$ has a linear relationship with the order of magnitude of $m_\phi$ as
\begin{align}
	\phi_{\rm crit} = 0.70 - 0.37 \log_{10}(m_\phi).
\end{align}

\begin{figure}
	\centering
	\includegraphics[width=0.42\textwidth]{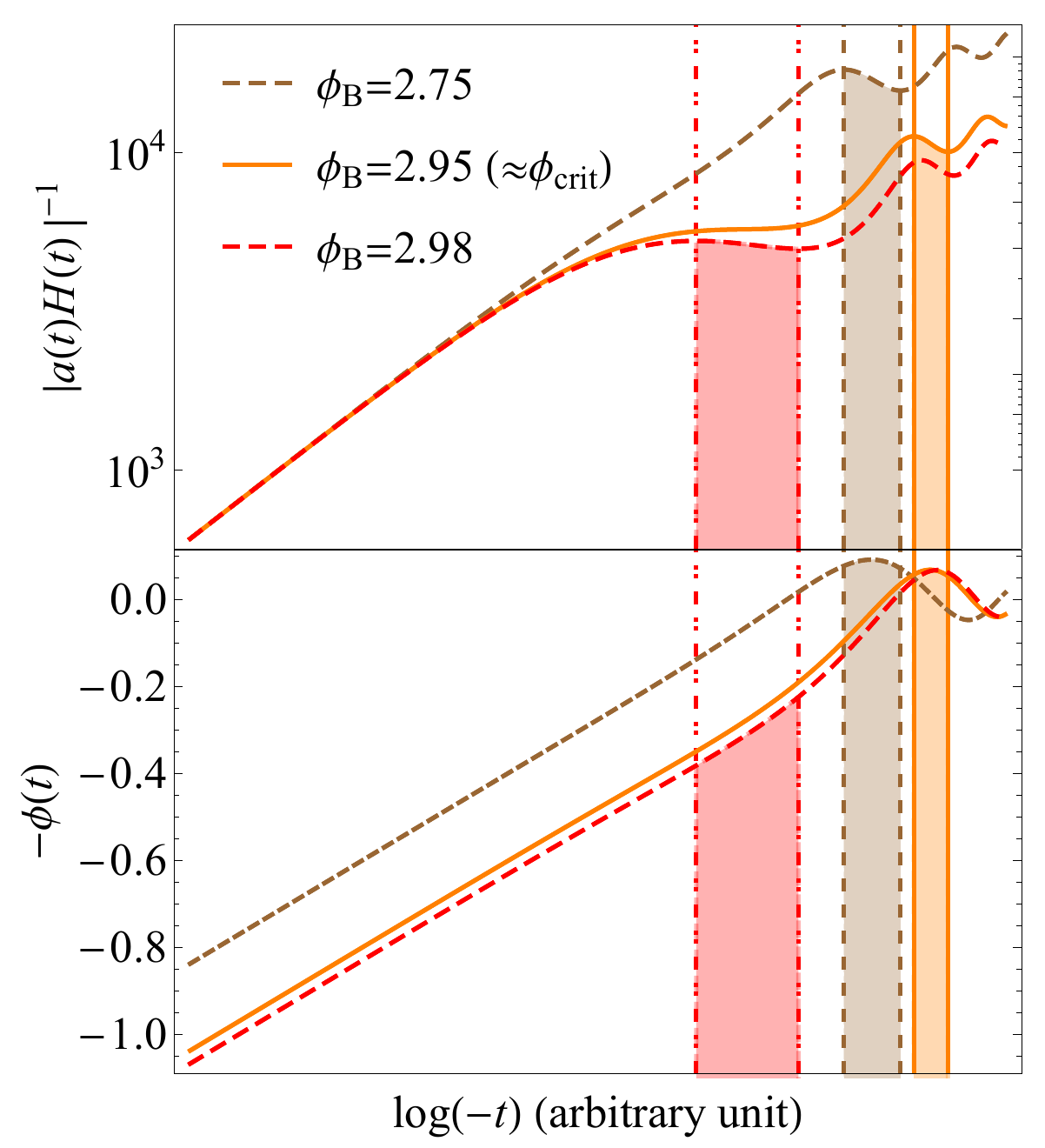}
	\caption{The time evolution of the comoving Hubble radius (upper panel) and the scalar field (lower panel) with $\phi_{\rm B}=2.75$ ($< \phi_{\rm crit}$; brown dashed), $2.95$ ($\approx \phi_{\rm crit}$; orange solid), and $2.98$ ($> \phi_{\rm crit}$; red dashed) for Chaotic potential. The vertical lines denote the beginning (right) and the end (left) of deflation, as the time goes leftwards in the plots.}
	\label{fig2.9}
\end{figure}

On the other hand, for each curve in Fig.~\ref{fig2.7}, we note that the value of $N_e^{\rm D}$ increases more dramatically when $\phi_{\rm B}$ departs from $\phi_{\rm crit}$ to a larger value than to a smaller value. This can be explained in Fig.~\ref{fig2.9} where we plot the comoving Hubble radius (upper panel) and the scalar field (lower panel) both as functions of time, for the case $m_\phi=10^{-6}$.
We consider three cases: $\phi_{\rm B} < \phi_{\rm crit}$ (brown dashed),
$\phi_{\rm B} \approx \phi_{\rm crit}$ (orange solid), and
$\phi_{\rm B} > \phi_{\rm crit}$ (red dashed).
In the upper panel, the parts of curves with negative slopes (increasing Hubble radius) indicate the periods when deflation takes place. These periods are shaded down to the lower panel and we see that the change in $\phi$ during deflation is obviously larger in the case when $\phi_{\rm B} > \phi_{\rm crit}$ (red dashed), resulting in the larger amount of deflation as seen in Fig.~\ref{fig2.7}.
We also note that in the lower panel of Fig.~\ref{fig2.9} the parts in the curves that cross $\phi=0$ can be thought of as the `inverse reheating', at which inflatons are produced by other particles. This is a period when all existing particles are converted to inflaton. This epoch always takes place before the deflation, so the scenario is like a mirror process of the inflation.

\begin{figure}
	\centering
	\includegraphics[width=0.48\textwidth]{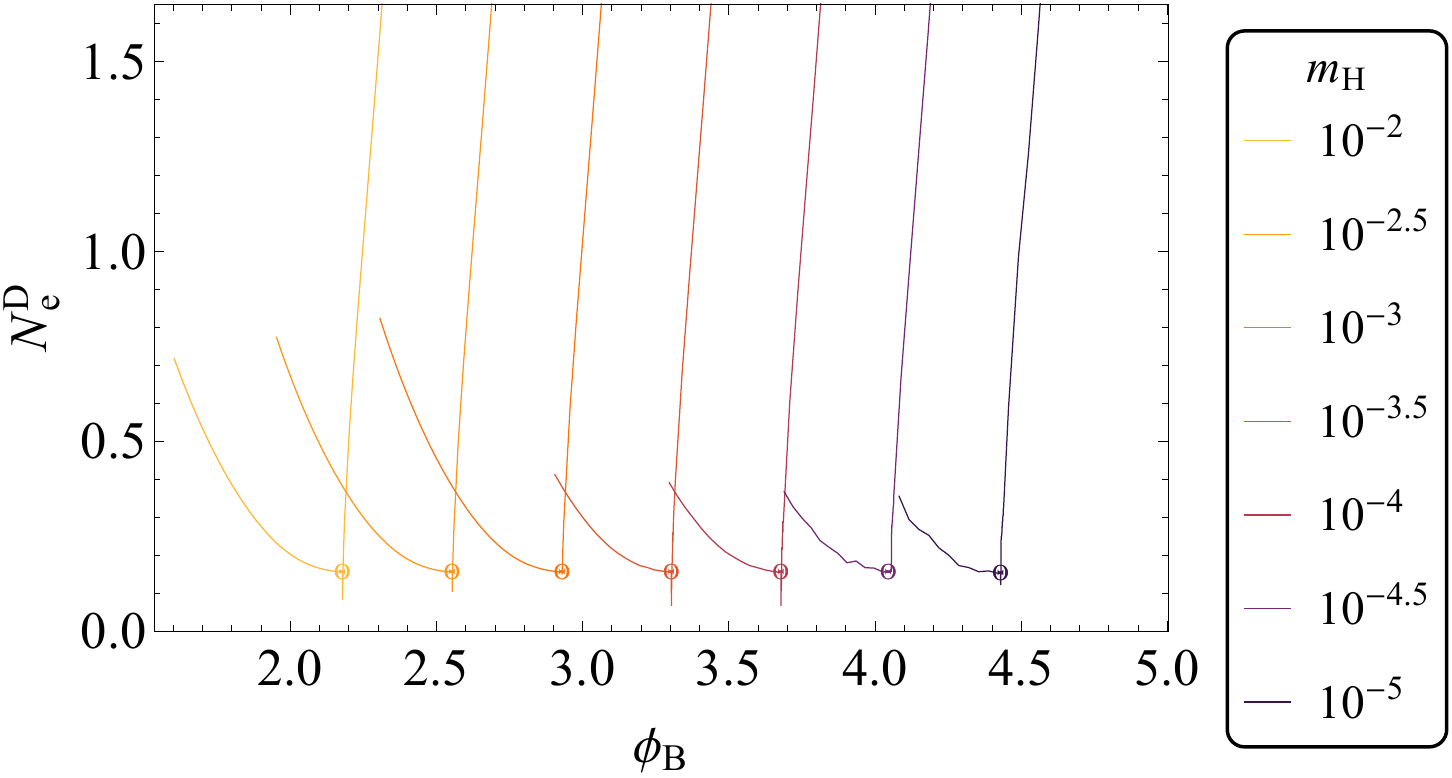}
	\caption{The number of $e$-foldings $N_e^{\rm D}$ for $R^2$ deflation as functions of $\phi_{\rm B}$ for different $m_{\rm H}$.}
\label{fig2.8}
\end{figure}

For the $R^2$ potential, the counter results are shown in Fig.~\ref{fig2.8} and Table~\ref{tab2.4}. There is a linear relationship between $\phi_{\rm crit}$ and the order of magnitude of $m_{\rm H}$ as well:
\begin{align}
	\phi_{\rm crit} = 0.68 - 0.75 \log_{10}(m_{\rm H}).
\end{align}
Again the minimum values of $N_e^{\rm D}$ in all cases are about the same, $0.15$, and for a fixed $m_{\rm H}$ the amount of deflation $N_e^{\rm D}$ increases more quickly when $\phi_{\rm B}$ departs from $\phi_{\rm crit}$ to a larger value than to a smaller value. We verified that the reason of this is the same as discussed in Fig.~\ref{fig2.9}.

%However, apart from these two selected inflationary models in this paper, there might be models that one cannot minimize deflation.
%To discuss this issue more generally we rewrite Eq.~\eqref{eq2.36}, by applying the slow-roll approximation, as
%\begin{align}
%	N_e^{\rm D} \simeq -8\pi \int_{\phi_{\rm e}^{\rm D}}^{\phi_{\rm b}^{\rm D}} \frac{V(\phi)}{V'(\phi)} d\phi = 8\pi \int_{\phi_{\rm b}^{\rm D}}^{\phi_{\rm e}^{\rm D}} n_e(\phi) d\phi,
%\end{align}
%where $V'(\phi)=dV/d\phi$ and $n_e(\phi)\equiv V(\phi)/V'(\phi)$.
%For a model with minimizable deflation one must acquire $n_e=0$ at a certain $\phi$ in the meantime $dn_e/d\phi>0$.
%We find that these two conditions can be reduced to a simple condition:
%\begin{align}
%	\exists~\phi\in\mathbb{R}: V(\phi)=0.
%\end{align}
%In other words, one is impossible to minimize deflation with a potential $V(\phi)$ which stays non-zero at all $\phi$.

In summary, for the selected inflatioary models in this paper, the amount of deflation is minimized when $\phi_{\rm B}$ reaches $\phi_{\rm crit}$.
For Chaotic potential this corresponds to the `most' shark-fin type (see Fig.~\ref{fig2.4}).
Since this $\phi_{\rm B}$ is model dependent and closely related to the amounts of deflation and inflation, we may use observations to confine $\phi_{\rm B}$ and thus the model parameters.

\begin{table}
	\vspace*{7mm}
	\centering
	\begin{tabular}{|m{2.55em}|m{2.55em}|m{2.55em}|m{2.55em}|m{2.55em}|m{2.55em}|m{2.55em}|m{2.55em}|m{0em}}
		\cline{1-8}
		$m_{\rm H}$ & $10^{-2}$ & $10^{-2.5}$ & $10^{-3}$ &	$10^{-3.5}$ & $10^{-4}$ & $10^{-4.5}$ & $10^{-5}$ & \\ [5pt]
		\cline{1-8}
		$\phi_{\rm crit}$ & $2.18$ & $2.55$ & $2.93$ & $3.30$ & $3.68$ & $4.05$ & $1.30$ & \\ [5pt]
		\cline{1-8}
	\end{tabular}
	\caption{The values of $\phi_{\rm crit}$ for different $m_{\rm H}$ in $R^2$ deflation. They correspond to the minima in Fig.~\ref{fig2.8}.}
	\label{tab2.4}
\end{table}

% -----------------------------------------------------------------------

\section{Conclusion}\label{sec6}

We employed the parameter $\phi_{\rm B}$ to discuss the time asymmetry in the cosmic background evolution with respect to the quantum bounce. 
It is particularly noted that the time-symmetric scenarios should be avoided because in such cases deflation and inflation may counteract each other, likely leaving the cosmological conundrums unresolved.
In the consideration of number of $e$-foldings,
there is a critical value of $\phi_{\rm B}$ at which the amount of deflation is minimized.
This critical value $\phi_{\rm crit}$ depends on the model parameters, namely the $m_\phi$ and $m_{\rm H}$ in the Chaotic and $R^2$ potentials respectively in our demonstrations.
Thus when we study any model in LQC, we should be cautious about the level of time asymmetry in order to have sufficient inflation that is not pre-canceled out by the deflation before the quantum bounce.
Within this context, other issues such as the cosmological perturbations also require proper treatment. In this regard we proposed a new formalism for evolving the tensor perturbations (gravitational waves) \cite{Chang2018a}. All these will need to pass the observational tests such as the cosmic microwave background in the near future.

% -----------------------------------------------------------------------

\acknowledgments
We acknowledge the support from the Ministry of Science and Technology, Taiwan (MOST 103-2628-M-002-006-MY4).

% -----------------------------------------------------------------------

\bibliography{reference_04}

\end{document}